 \scrollmode
 \documentstyle[12pt,A4]{article}
\begin{document}
\title{{\bf
On the random filling of $R^d$ by non-\-overlapping d-dimensional cubes}}
 \author{{\bf B. Bonnier, M. Hontebeyrie, and C. Meyers  }\\
{\em Laboratoire de Physique Th\'eorique,$^{\dag}$} \\
{\em Universit\'e de Bordeaux I, }\\
{\em 19 rue du Solarium, F-33175 Gradignan Cedex}}
\date{}
\begin{titlepage}
\maketitle
\thispagestyle{empty}
\vspace{2 cm}
\begin{abstract}
We compute the time-dependent coverage in the random sequential
adsorption of aligned d-dimensional cubes in $R^d$ using time-series
expansions.
The seventh-order series in 2, 3 and 4 dimensions is resummed in order to
predict the coverage at jamming. The result is in agreement with
Monte-Carlo simulations.
A simple argument, based on a property of the perturbative expansion
valid at arbitrary orders, allows us to analytically derive some
generalizations of the Pal\'asti approximation.
\end{abstract}
\vfill
LPTB 93-3\\
February 1993
BITNET Addresses:
{\bf Bonnier@FRCPN11}, {\bf Beyrie@FRCPN11},
{\bf Meyers@FRCPN11}\\
$\dag$ $\overline{\mbox{Unit\'e associ\'ee au CNRS}}$, U.A.764\\
\end{titlepage}

\section{Introduction}
{}~

    Random sequential adsorption (RSA) is a model for the random adsorption of
non-overlapping objects on a surface. RSA is a conceptually simple
model for irreversible behaviour. It has a rich variety of forms
which are explained, for example, in the review paper of J.W.
Evans~\cite{Ermp}.
Here, we consider the random filling of $R^d$ by non-overlapping aligned
d-dimensional cubes. This is a simple model, whose one-dimensional version
has been  solved since a long time \cite{old}.
In comparison, the level of understanding of this model in higher dimensions
is low.
It has been studied through
2, 3 and 4 dimensional Monte-Carlo (MC) simulations and through a 2-d
perturbative analysis \cite{DWJ}.
A density series expansion \cite{TST} can also be used to analyze
this model. It is equivalent to a direct perturbative expansion of the
coverage in the time variable.

    A striking particularity of the model concerns the
 behavior of the fraction of space, $\theta_d$, filled at jamming.
$\theta_d$ is numerically found to be close to the jamming coverage in
one dimension raised to the d-th power. The relation
$\theta_d = \theta_1^{~d}$,
proposed long ago by Pal\'asti~\cite{Palasti},
is, in fact, one of the motivations
of some MC experiments with an increasing precision.
Nowadays,
 they allow us
to rule out this exact	relation. Nevertheless, as an approximation it appears
to have a puzzling accuracy, e.g. of the order of 0.5\% in two
dimensions \cite{BZV,PWN}.

The Pal\'asti relation is one of our main interests.
 More generally, we undertake the reconstruction of the coverage $\theta_d(T)$
 in 2, 3 and 4-d using as basic input its perturbative expansion (PE) in the
 time variable $T$.

In the next section, this expansion is derived up to seventh order
by using diagramatic  methods already given in Ref.\cite{DWJ,E}.
 The complexity of this expansion, even at moderate orders (6th or 7th),
 necessitates a computer algorithm. We will only give the results in 2, 3 and
4-
(because the full analytical result in d is too long).

Using standard approximation methods, which combine PE and asymptotic
 ($T \rightarrow \infty  $) expansions, we extend the analysis of $\theta_d$
of Ref.\cite{DWJ} to 3 and 4 dimensions.
This analysis is done in section 2, where it  is found
that $\theta_d \approx \theta_1^{~d}$ with some decreasing
 precision as	$d$ increases.
 In section 3, we provide a functional understanding of the Pal\'asti
 relation as a
 consequence of a    particular property of the PE, valid at arbitrary orders.
 Finally, by taking into account the asymptotic regime, we give some improved
 generalizations of the Pal\'asti approximation.

 \section{The d-dimensional perturbative expansion}
{}~

We will study  the  RSA of hypercubes in the continuum limit on $R^d$.
This is  defined as the scaling limit of RSA of cubes on a d-dimensional
hypercubical
lattice. The discrete model, in which the size of the adsorbed cubes enters as
a parameter, has its own interest and it has been also investigated
\cite{old,DWJ,PWN,NE} in 1 and 2-dimensions.
At small sizes, the Pal\'asti approximaton works perfectly, and more generally,
the model is easier to handle \cite{DWJ,STRR}. We consider here the
discretisation as a convenient regularisation of the PE from which we will
keep, order by order, the scaling limit.

We follow the definition of the discrete model given in Ref.\cite{PWN}.
The adsorbed cubes have their edges parallel to the $d$ axes of the lattice,
and they are made of $k^d$ unit cells, $k$ being an integer.
 Their volume $l^d = (k b)^d$ enters in the
definition of the dimensionless time, $T$, $T= \rho t l^d$, in which $b$
is the lattice spacing.
$\rho$ is the  attempt rate to adsorb cubes per volume unit.
On the lattice, the number of attempts per site after time $t$ is
thus $\rho t b^d$, i.e. $T/k^d$. Therefore the coverage can
be written as
     \begin{equation}
\theta_d(k,T)  = k^d~\rho_d(k,\frac{T}{k^d})	     \label{1}
 \end{equation}
The coverage $\theta_d(T)$ in the continuum limit is then given by the
limit of $\theta_d(k,T)$ as $k$ goes to infinity with $T$ fixed.

The PE of $\rho_d(k,t)$ is:
\begin{equation}
\rho_d(k,t) = t - \sum_{n \geq 2} C_n(k,d)~\frac{(-t)^n}{n!}	  \label{2}
\end{equation}
where the first order in time is $t$, because the first
adsorption attempt on an empty lattice is always accepted.
We refer to Ref.\cite{DWJ} for a comprehensive and complete derivation of the
PE. It starts with an operator formalism of the RSA mechanism and ends up
with diagrammatic rules, that are
sketched below, for computing the coefficients $C_n(k,d)$.

To generate the n-th order diagrams, one expands the  expression:

$$ \prod_{i=2}^n ( 1 - \prod_{j=1}^{i-1} ( 1 - K_{i~j})  )  $$
The set
of variables defining a deposition is denoted by $i$ , and  $K_{i~j}$
is a hard Mayer function ($K_{i~j}=1$ only if $i$ and $j$ are
overlapping objects, $0$ otherwise). All monomials of this polynomial are
connected
labeled graphs which are regrouped in classes of the same topology.
Thus, they appear with combinatorial weigths $w_{n,i}$ in such a way that
\begin{equation}
C_n(k,d) = \sum_{i} w_{n,i}~ I(\Gamma_{n,i})   \mbox{~~.}  \label{Cn}
\end{equation}
$I(\Gamma_{n,i})$ is the graph contribution which embodies the
details of the model.

 In order to compute $I(\Gamma_{n,i})$, one has to do the summation over
 $\{ x_l \}$, the n vertices of $\Gamma_{n,i}$. At this stage,
the vertices, $\{ x_l \}$, are to be understood as
 d-dimensional lattice points, i.e. $x_l \equiv (x_l^1, x_l^2, \dots ,
x_l^d)$, because a d-dimensional aligned hypercube can be characterized by the
position of a distinguished point, for example a corner.  Finally, the graph
contribution is  \begin{equation}
 I(\Gamma_{n,i}) = \sum_{\{ x \}}~\prod_p K(x_{l_p},x_{m_p}) \equiv
I_{n,i}(k,d) \mbox{~~.}  \label{graph} \end{equation}
The product $\prod_i$ is done on the links of $\Gamma_{n,i}$ and the
 sum is over all the lattice degrees of freedom for every variable $x_l$,
except	one which can be frozen at the origin by transational invariance.

 Any graph contribution, $I_{n,i}(k,d)$, is thus the $d$-th power of the
 contribution of the same graph in one dimension.
 \begin{equation}
 I_{n,i}(k,d)	 = I_{n,i}(k,1)^d	\mbox{~~.}		 \label{Intd}
 \end{equation}
This simple property, that we shall use in the next section, is a
consequence of the construction rule (\ref{graph}) and of the factorization
property
of hard Mayer functions for hypercubes: $K_d(x,y)=\prod_{i=1}^d K_1(x^i,y^i)$.

Finally, in the continuum limit, the PE of $\theta_d(T)$ is,
 from (\ref{1}) and (\ref{2}):

\begin{equation}
\theta_d(T) = T - \sum_{n \geq 2} C_n(d)~\frac{(-1)^n}{n!}~T^n	  \label{PE}
\end{equation}
where, from (\ref{Cn}-\ref{Intd})
\begin{equation}
C_n(d) =  \sum_{i} w_{n,i} \lim_{k \rightarrow \infty}
\{ k^{1-n} I_{n,i}(k,1) \}^d	= \sum_{i} w_{n,i}~I_{n,i}^{~~d}    \label{Id}
\end{equation}
The coefficients $C_n(d)$ are well defined, because $I_{n,i}(k,1)$ are
polynomials of degree $(k-1)$ in the variable $k$ that we have analytically
computed. Their leading coefficients are denoted by $I_{n,i}$ in
Eq.(\ref{Id}).

We have thus, stored in our computer algorithm, all the $I_{n,i}$ and
$w_{n,i}$ necessary to know $C_n(d)$ analytically in $d$, up to the seventh
order in $T$. Nevertheless we do not explicitly write them  here due to their
number
(for $n = 7$ there are about $10^3$ graphs). At fixed $d$, $C_n(d)$ are
just the rational numbers that we give in Table I for $d=2,3$ and $4$. We have
performed various tests of our expansion. The simplest is for $d=0$ where
$\theta_0(T) = 1 - e^{-T} $. For $d=1$ the exact solution is known for any $k$
and in the scaling limit $\theta_1(T)$ reads:
\begin{equation}
\theta_1(T) = \int_0^T dt e^{-2 g(t)}
\mbox{~~~~    with~~~~ } g(t)=\int_0^t dx \frac{1-e^{-x}}{x} \quad .
\end{equation}
The agreement was checked analytically in $k$, including the scaling limit.
Finally, in 2-d, we recover the expansions given for $k=2$ and $k=\infty$
in Ref.\cite{DWJ}.

These coefficients $C_n(d)$ can be used in various ways to approximate
$\theta_d(T)$. We give, in the following, our analysis.
In addition to the PE Eq.(\ref{PE}), one knows how the jamming limit,
 $\theta_d = \theta_d(T=\infty)$, is approached \cite{S,PWN}.
\begin{equation}
\theta_d(T) - \theta_d = O[ \ln{T}^{d-1} / T ]	  \label{GT}
\end{equation}
Standard methods which embody these two kinds of information have shown, in
other contexts, their ability to reasonably approximate the coverage even
in its asymptotic regime \mbox{\cite{DWJ,E}.}

We give, as a first attempt, the predictions for $\theta_d$ from a
staigthforward extension of the method used in Ref.\cite{DWJ}. One uses the
mapping $\omega(T)$,
\begin{equation}
\omega(T)=1-\frac{1+\ln{(1+\frac{b-1}{d-1}T)}^{d-1}}{1+ bT}
\mbox{~~~~~, $b \geq 1$}
 \end{equation}
which behaves as $T$ for small $T$ and as $1-\ln{T}^{d-1}/ b T$
for $ T \rightarrow \infty$, to convert the PE, Eq.(\ref{PE}), into a power
series
of $\omega$. The Pad\'e approximants formed from this series, when evaluated
at $\omega=1$, give a $\theta_d$ consistent with the behaviour
(\ref{GT}), for any choice of the $b$ which is used as a variational parameter.

Let us first recall the  2-d results from Ref.\cite{DWJ}. Among the sixth and
seventh order approximants, [5,2], [3,4], [4,2] and [3,3] cross near $b=1.30$
yielding $\theta_2 = 0.52625(5)$, in agreement with the data of
Ref.\cite{BZV}, $\theta_2 = 0.562009(4)$.
As $d$ increases, we expect some continuity, and we
indeed observe a crossing at $b=1.40$ for the [4,2] and [3,4] approximants,
and for the remaining ones [5,2] and [3,3] there is a crossing at $b=1.45$.
Nevertheless, the prediction, now, loses its accuracy, because these
intersections give
$\theta_3 = 0.40$ and $0.46$, respectively. Moreover, in 4-d this intersections
disappear and the best determination of $b$ is $b \simeq 2.3-2.5$.
At this value, the
[4,3], [4,2], [3,3] and [3,2] Pad\'e-approximants give $\theta_4 = 0.30(1)$.

These values are consistent with the MC data since $\theta_3 = 0.4227(6)$,
$0.4262$ , $0.430(8)$ and $0.422(8)$ according to the
Refs.\cite{JT,BS,C,N} respectively and $\theta_4 = 0.3129$ or $0.3341$ from
Ref.\cite{JT} and Ref.\cite{BS} respectively. Obviously, the Pal\'asti values,
$\theta_1^d=0.7476$, $0.5589$, $0.4178$, and $0.3124$, $d=2, 3, 4$ cannot
be discussed (except in 2-d) within the poor precision of the previous results.
Nevertheless, we show in the following section that perturbation theory is able
to provide a functional insight into the Pal\'asti approximation. It goes
beyond a numerical coincidence.

\section{Variational approach to the Pal\'asti approximation}
{}~

      The  RSA mechanism of deposition of aligned cubes is
characterized
by  power law behaviors in $d$ for both the coverage at jamming, in the
Pal\'asti
approximation, and the graph-integrals, in the framework of the perturbative
expansion.
We shall give in this section
a variational derivation of this puzzling approximation together with an
extension to finite values of $T$ .

Starting with the representation of $\theta_d(T)$ given by
Eqs.(\ref{PE},\ref{Id}),
\begin{equation}
\theta_d(T) = \sum_{i,~n\geq 1} \Lambda_{i,n} ~ I_{i,n}^{~~d}~~T^n
\label{PEd} \end{equation}
we write, in order to emphasize the role of the power law behavior
(\ref{Intd}), that

\begin{equation}
I_{i,n}^{~d} = (I_{i,n}-a b^n)^d+
\sum_{r=0}^{d-1} (-1)^{d-r+1} \frac{d!}{r! (d-r)!}~ a^{d-r}~ b^{n(d-r)}  ~~
I_{i,n}^{~~r}
\label{bin} \end{equation}
This is nothing but a binomial identity,
valid for any $a$ and $b$. We
can insert (\ref{bin}) into (\ref{PEd}) to obtain a two-component
representation of $\theta_d(T)$:
\begin{equation}
\theta_d(T) = \bar{\theta_d}(T)   +\epsilon_d(T)	 \label{theta}
\end{equation}
where
\begin{equation}
\bar{\theta_d}(T)=
\sum_{r=0}^{d-1} (-1)^{d-r+1} \frac{d!}{r! (d-r)!} a^{d-r}~~
\theta_r(b^{d-r} T )
\label{thetab} \end{equation}
\begin{equation}
\epsilon_d(T) = \sum_{i,~n \geq 1} \Lambda_{i,n}~~ ( I_{i,n} - a b^n )^d~~ T^n
 \label{eps} \end{equation}

In the representation (\ref{theta}-\ref{thetab}), $\bar{\theta_d}(T)$ involves
all the lower dimensionality coverages, and the a priori $d$- and $T$-
dependent parameters $a$, $b$.
 A natural way to choose of $a$ and $b$
is thus to look for maxima of $\bar{\theta_d}(T)$, i.e. minima of the
perturbative component $\epsilon_d(T)$.

Beginning with the 2-d case, one finds that
$\bar{\theta_2}(T)= 2 a \theta_1(b T) - a^2 \theta_0(b^2 T)$,
has a maximum with respect to $a$ :
$$	 \theta_2^*(T)= \frac{	\theta_1^2(b T) }{ \theta_0(b^2 T)}
 \label{pal2} $$
for $a= \theta_1(b T)/\theta_0(b^2 T)$, i.e. a value of $a$ where $\epsilon_1$
vanishes. If $\epsilon_2$ is $0$  , this is nothing but the Pal\'asti
approximation at $T=\infty$.

 This result can be extended to any $d$:
Under the assumption $\epsilon_r = 0$ for $1\leq r \leq d-1$,
$\bar{\theta}_d(T)$ has an extremum $\theta_d^*(T)$:
$$
\theta_d^*(T)= \frac{  \theta_1^d(b^{d-1}T) }{ \theta_0^{d-1}(b^d T)}~
=~  \frac{  \theta_{d-1}^2(b T) }{ \theta_{d-2}(b^2 T)}
 \label{pald} $$
for $a= \theta_1(b^{d-1}T)/ \theta_0(b^dT)$.

 This is simply because the minimizing
condition can be reexpressed in terms of $\epsilon_{d-1}$ if the $a$
parameter is assumed to be $d$-independent. The relation is:

$$\frac{\partial \epsilon_{d}(T) }{\partial a}= - d~ \epsilon_{d-1}(bT)
\mbox{~~.}$$
We are thus lead to write the PE (\ref{theta}) of $\theta_d(T)$ under the
form:
\begin{equation}
\theta_d(T) = \theta_d^*(T)   +\epsilon_d(T)	     \label{thetam}
\end{equation}
where we choose
\begin{equation}
\theta_d^*(T) =  \frac{  \theta_{d-1}^2(b T) }{ \theta_{d-2}(b^2 T)}
 \label{paldm} \end{equation}
as a convenient generalization of the Pal\'asti approximation on the whole
time-range. To go further, one can eliminate $T$, order by order, in terms
of $\theta_d^*(T)$     ($\theta_d^*(T)~\sim~T$	at
$T=0$ and		 $\theta_d^*(T)$ increases  if~$b\geq~1$), and
then, substitute this expression in $\epsilon_d(T)$, ($\epsilon_d(T)~\sim~T^2
$). One finally obtains $\theta_d(T)$ as a power series in
$\theta_d^*(T) $, i.e. as a perturbative expansion where the first
order is the Pal\'asti approximation.

In this approach, it remains to take into account the asymptotic behaviour
(\ref{GT}). One oberves that
$\frac{d}{dT}( T \theta_d(T))$ and $\theta_d^*(T) $
reach their asymptotic limit in the same way, $ O[ \ln{T}^{d-2} / T ]  $.
On the other hand,
$\frac{d}{dT}( T \theta_d(T))$
is perturbatively of the form (\ref{PEd}), so that the previous arguments
apply. Thus we propose:
$$
\theta_d(T) = \frac{1}{T}\int_0^T dT \{ \theta_d^*(t) +\epsilon_d(t) \}
     \label{thetaex} $$
as an approximation of	$\theta_d(T)$ with a minimal component $\epsilon_d(T)$.
The jamming coverage is then
$\theta_d = \theta_d^*(\infty) + \epsilon_d(\infty)$ ,	where\\
$$\theta_3^*(\infty) = \theta_2^2 / \theta_1 ~=~ 0.4225 ~$$
$$\theta_4^*(\infty) = \theta_3^* / \theta_2 ~=~ 0.3176 .$$
We have to verify, in PE, that $\epsilon_d(T)$ is small and that a precise
determination
of the $b$ parameter can be done. We express $\epsilon_d(T)$ as a power series
in $\theta_d^*(T)$ and form its Pad\'e table. We find
$\epsilon_d(T=\infty)$	from these approximants when
$\theta_d^*=\theta_d^*(\infty)$. Here, $b$ is used as a variational
parameter.
\newpage
In 2-d, 3-d and 4-d, we find a minimal dispersion of the
[3,4], [4,3], [5,2], [4,2], [3,3], [2,4], [2,3] and [3,2] Pad\'es at the
values:\\ \begin{center}
\begin{tabular}{ccc}
$d=2$ & $b \simeq 1.32~~~ $ & $ \epsilon_2(\infty) = 0.004(2)$~~\\
$d=3$ & $b =	  1.35(3) $ & $ \epsilon_3(\infty) = 0.00 (1)$~~  \\
$d=4$ & $b =	  1.40(5) $ & $ \epsilon_4(\infty) = 0.01 (1)$~.\\
\end{tabular}
\end{center}
$\epsilon_d(T)$ is found to be positive on the whole physical region
(except near $T=\infty$ for some marginal solutions in 3-d)
and is of the order of $\epsilon_d(\infty)$. We finally obtain that
$$
\theta_3 = 0.42(1)  \mbox{~~ and ~~}  \theta_4 \simeq 0.32(1) \mbox{~.}
$$
In most of the cases, the positivity of $\epsilon_d(T)$ indicates
the violation of the Pal\'asti conjecture.

\section{Conclusion}
{}~

This study has been devoted to the RSA coverage of $R^d$ by $d$-dimensional
aligned hypercubes, especially in 3 and 4 dimensions.
For these dimensions, we have explicitely presented the perturbation result
to seventh-order. The series in other dimensions
and any discrete case are available upon request. The continuum
model at long times is expected to be more
difficult to describe than its discrete analogue. This is supported by the
fact that we cannot predict $\theta_d$ beyond a moderate accuracy with
our perturbative information. Nevertheless, we indicate, in this
framework, how one can approach the Pal\'asti conjecture and its corrections.

To conclude, we want to mention that the simple rule
$\theta_d \sim \theta_{d-1}^{~~2}/ \theta_{d-2} $
applies to various discrete models, in particular RSA on $d$-dimensional
lattice with nearest-neighbour exclusion  \cite{BK1},
where $\theta_0 = 1/2$ and $ \theta_1 = (1-e^{-2})/2 $. As $d$ runs from
2 to 6, we obtain the sequence of approximations
(0.374, 0.307, 0.254, 0.229, 0.206) which are
quite close to the corresponding data
(0.364, 0.304, 0.264, 0.233, 0.209).

\newpage
\begin{center}
\newcommand{\ds}{$\displaystyle}
\begin{tabular}{|c|c|c|c|} \hline
\multicolumn{4}{|c|}{\ds C_n(d) $} \\
\hline
 n &	  d=2	   &	d=3	&     d=4    \\
\hline
  & & & \\
2 & $4$   &  $ 8$ & $ 16 $     \\
  & & & \\
3 &$ 23$  &  $ 101$ & $ 431$   \\
  & & & \\
4 & $168 $	& \ds\frac{45920}{27}$& \ds\frac{ 432032}{27 } $\\
  & & & \\
5 &\ds\frac{105895}{72} $ & \ds\frac{30622397}{864}$  &
\ds\frac{7785588283}{10368} $ \\
  & & & \\
6 &\ds\frac{6709687}{450}$ & \ds\frac{11820085427}{13500}$ &
\ds\frac{17179096575091}{405000}$    \\
  & & & \\
7 &\ds\frac{1385692277}{8100}$ & \ds\frac{5815177641671}{233280}$ &
\ds\frac{183314989218969359}{65610000}$   \\
  & & & \\
\hline
\hline
\end{tabular}

\vspace{3cm}
{\bf Table I: }~Coefficients of the time-series expansion of
the coverage $ \theta_d(T)$
defined  by
$$\theta_d(T) = T - \sum_{n\geq 2} \frac{(-1)^n}{n!}~ C_n(d)~ T^n\mbox{~.}$$\\
\vfill
\end{center}

\newpage

\begin{thebibliography}{99}
\bibitem{Ermp}
J.W. Evans,   Review of Modern Physics (to be published).
\bibitem{old}
P.J. Flory, J. Am. Chem. Soc. {\bf 61}, 1518, (1939).\\
J.J. Gonzalez, P.C. Hemmer and J.S. Hoye, Chem. Phys. {\bf 3}, 228, (1974).\\
A. R\`enyi, Sel. Transl. Math. Stat. Prob. {\bf 4}, 203, (1963).
\bibitem{DWJ}
R. Dickman, J.S. Wang and J. Jensen, J. Chem. Phys. {\bf 94}(12), 8252, (1991).
\bibitem{TST}
G. Tarjus, P. Schaaf and J. Talbot, J. Stat. Phys. {\bf 63}, 167, (1991).
\bibitem{Palasti}
I. Pal\'asti, Publ. Math. Inst. Hung. Acad. Sci. {\bf 5}, 353, (1960).
\bibitem{BZV}
B.J. Brosilov, R.M. Ziff and R.D. Virgil, Phys. Rev. A {\bf 43}, 631, (1991).
\bibitem{PWN}
V. Privman, J.S. Wang and P. Nielaba, Phys. Rev. B {\bf 43}, 3336, (1991).
\bibitem{E}
J.W. Evans, Phys. Rev. Lett. {\bf 62}, 2642 (1989).
\bibitem{NE}
R.S. Nord and J.W. Evans, J. Chem. Phys. {\bf 82} (6), 2795, (1985).
\bibitem{STRR}
P. Schaaf, J. Talbot, H.M. Rabeony and H. Reiss,
J. Chem. Phys. {\bf 92}, 4826, (1988).
\bibitem{S}
R.H. Swendsen, Phys. Rev. A {\bf 24}, 504, (1981).
\bibitem{JT}
W.S. Jodrey and E.M. Tory, J. Stat. Comp. Sim. {\bf 10}, 87, (1980).
\bibitem{BS}
B. Blaisdell and H. Solomon, J. Appl. Prob. {\bf 19}, 382, (1982).
\bibitem{C}
D.W. Cooper, J. Appl. Prob. {\bf 26}, 664, (1988).
\bibitem{N}
R.S. Nord, J. Stat. Comp. Sim. {\bf 39}, 231, (1991).
\bibitem{BK1}
A. Baram and D. Kutasov, J. Phys. A: Math. Gen.  {\bf 22}, (L251), 1989.
\end{thebibliography}
\end{document}